\documentclass{pasj00}
\twocolumn
\newcommand{\degg}{\hbox{$^\circ$}}
\newcommand{\arcm}{\hbox{$^\prime$}}

\begin{document}
\SetRunningHead{Reeves et al.}{Revealing the High Energy Emission From MCG\,-5-23-16}
\Received{2006/07/27}
\Accepted{2006/09/18}

\title{Revealing the High Energy Emission from the Obscured Seyfert Galaxy 
MCG\,-5-23-16 with {\it Suzaku}}



%

\author{James N. \textsc{Reeves}\altaffilmark{1,2}
Hisamitsu \textsc{Awaki}\altaffilmark{3}
Gulab C. \textsc{Dewangan}\altaffilmark{4}
Andy C. \textsc{Fabian}\altaffilmark{5} 
Yasushi \textsc{Fukazawa}\altaffilmark{6}\\
Luigi \textsc{Gallo}\altaffilmark{7,8}
Richard \textsc{Griffiths}\altaffilmark{4}
Hajime  \textsc{Inoue}\altaffilmark{8} 
Hideyo  \textsc{Kunieda}\altaffilmark{8,9}  
Alex \textsc{Markowitz}\altaffilmark{1} \\
Giovanni \textsc{Miniutti}\altaffilmark{5}
Tsunefumi  \textsc{Mizuno}\altaffilmark{6}
Richard \textsc{Mushotzky}\altaffilmark{1}
Takashi \textsc{Okajima}\altaffilmark{1,2}
Andy \textsc{Ptak}\altaffilmark{1,2} \\
Tadayuki \textsc{Takahashi}\altaffilmark{8}
Yuichi \textsc{Terashima}\altaffilmark{3,8}
Masayoshi \textsc{Ushio}\altaffilmark{8} 
Shin  \textsc{Watanabe}\altaffilmark{8} 
Tomonori \textsc{Yamasaki}\altaffilmark{6} \\ 
Makoto \textsc{Yamauchi}\altaffilmark{10} 
and Tahir \textsc{Yaqoob}\altaffilmark{1,2}}

\altaffiltext{1}{Exploration of the Universe Division, Code 662, NASA Goddard Space 
Flight Center, Greenbelt Road, \\ Greenbelt, MD 20771, USA}
\email{jnr@milkyway.gsfc.nasa.gov}
\altaffiltext{2}{Department of Physics and Astronomy, Johns Hopkins University,  
3400 N Charles Street, \\ Baltimore, MD 21218, USA}
\altaffiltext{3}{Department of Physics, Ehime University, 
Matsuyama 790-8577, Japan}
\altaffiltext{4}{Department of Physics, Carnegie Mellon University, 5000 Forbes Avenue,  
Pittsburgh, \\ PA 15213, USA}
\altaffiltext{5}{Institute of Astronomy, University of Cambridge, Madingley Road, 
Cambridge, \\ CB3 0HA, UK}
\altaffiltext{6}{Department of Physics, Hiroshima University, 1-3-1 Kagamiyama, 
Higashi-Hiroshima, \\ Hiroshima 739-8526, Japan}
\altaffiltext{7}{Max-Planck-Institut f\"ur extraterrestrische Physik, Postfach 1312, 
Garching, Germany}
\altaffiltext{8}{Institute of Space and Astronautical Science, Japan Aerospace Exploration 
Agency, Yoshinodai 3-1-1, \\ Sagamihara, Kanagawa 229-8510, Japan}
\altaffiltext{9}{Department of Physics, Nagoya University, Furo--cho, Chikusa, 
Nagoya 464-8602, Japan} 
\altaffiltext{10}{Department of Applied Physics, University of Miyazaki, 
1-1, Gakuen-Kibanadai-Nishi, \\ Miyazaki 889-2192, Japan}
\KeyWords{galaxies: individual (MCG\,-5-23-16);  galaxies: active; galaxies: Seyfert; 
X-rays: galaxies} 

\maketitle

\begin{abstract}

We report on a 100\,ks Suzaku observation of the bright, 
nearby ($z=0.008486$) Seyfert 1.9 galaxy MCG\,-5-23-16. 
The broad-band (0.4--100\,keV) X-ray spectrum allows us to determine the 
nature of the high energy emission with little ambiguity. 
The X-ray continuum consists of a cutoff power-law of 
photon index $\Gamma=1.9$,   
absorbed through Compton-thin matter of column density 
$N_{\rm H}=1.6\times10^{22}$\,cm$^{-2}$. A soft excess is observed below 
1\,keV and is likely a combination of emission from scattered continuum photons 
and distant photoionized gas. The iron K line 
profile is complex, showing narrow neutral 
iron K$\alpha$ and K$\beta$ emission, as well as a broad line 
which can be modeled by a moderately inclined 
accretion disk. The line profile shows either the disk 
is truncated at a few tens of gravitational radii, 
or the disk emissivity profile is relatively flat.  
A strong Compton reflection component is detected above 10\,keV, 
which is best modeled by a combination of 
reflection off distant matter and the accretion disk. 
The reflection component does not appear to vary. The overall picture 
is that this Seyfert\,1.9 galaxy is viewed at moderate ($\sim50$\degg) 
inclination through Compton-thin matter at the edge of a Compton-thick torus 
covering $\sim2\pi$\,steradians, consistent with unified models. 
\end{abstract}

\section{Introduction}

Determining the origin of the iron K emission line is one of the
fundamental issues in high energy research on Active Galactic Nuclei, as
it is regarded as the most direct probe available of the inner accretion
disk and black hole. Indeed the iron line diagnostic in AGN
first became important during the Ginga era, showing
that the 6.4 keV iron K$\alpha$ emission line was common amongst Seyfert
galaxies \citep{Pounds1990}. The associated 
reflection hump above 10 keV, produced by Compton down-scattering 
of higher energy photons, showed the iron line emission arises from 
Compton-thick material, possibly the accretion disk \citep{LW88, George91}. 
The higher (CCD) resolution spectra available with ASCA appeared to
indicate that the iron line profiles were broad and asymmetrically skewed (to
lower energies), which was interpreted as evidence
that the majority of the line emission
originated from the inner accretion disk around the massive black hole
\citep{Tanaka95, Nandra97, Reynolds97}.

The picture emerging from the study of the iron K line with 
XMM-Newton and Chandra is much more complex. The presence of 
a narrower 6.4 keV iron emission component, from more distant matter 
(e.g. the outer disk, BLR or the molecular 
torus) is commonplace in many type I AGN 
\citep{YP04, Page04}. 
In contrast the broad, relativistic component of the iron line profile 
appears to be weaker than anticipated 
and in some cases may be absent altogether, 
e.g. NGC 5548 \citep{Pounds03a}; NGC 4151 \citep{Schurch03}. 

Furthermore, it is possible that complex absorption in some objects
may also effect the modeling of the iron K-line and reduce its strength; e.g. 
in NGC 3783 \citep{R03} and NGC 3516 \citep{Turner05}. 
The possible presence of the reflection component hardening the spectrum towards higher 
energies \citep{George91} can also complicate fitting the broad iron 
K line, if bandpass above 10\,keV is not available. 
This is where new observations with Suzaku can provide an important 
breakthrough, by determining the underlying AGN continuum 
emission over a wide bandpass (e.g. from 0.3 keV to $>100$\,keV), 
thereby resolving the ambiguities present in fitting the iron K-shell band.

MCG\,-5-23-16 is an X-ray bright, nearby ($z=0.008486$ or 36\,Mpc, \cite{Wegner03}) 
AGN that is classed optically as a Seyfert 1.9 galaxy \citep{Veron80} 
and is known to possess moderate absorption in the soft X-ray band 
\citep{Mushy82}. It is one of the brightest Seyfert
galaxies in the 2-10 keV band, with a historically typical flux of
$7-10\times10^{-11}$~erg~cm$^{-2}$~s$^{-1}$ \citep{MW04},
which makes it an ideal object to study
in the iron K band and at high energies to measure any Compton reflection 
component. Indeed of the objects that now show
evidence for a broad redshifted
iron K line, MCG\,-5-23-16 appears to be one of the
more robust examples from XMM-Newton data 
\citep{Dewangan03, Balestra04} or with ASCA \citep{Weaver97}.  

Here we report on a 100\,ks observation of 
MCG\,-5-23-16, performed in December 2005 with Suzaku \citep{Mitsuda06}. 
In Section 2, the Suzaku observations of MCG\,-5-23-16 are outlined, 
whilst in 
Section 3 the detailed modeling of the time-averaged spectrum
is performed. In Section 4 the variability of the iron line and reflection 
component are compared while in Section 5, modeling of the reflection 
component is discussed.

\begin{figure}
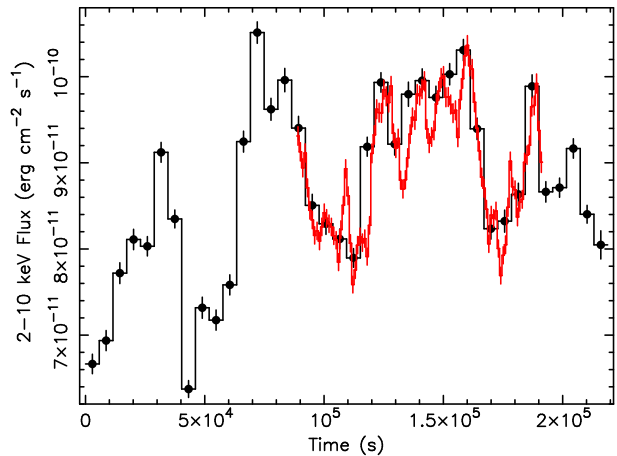

\begin{center}
\rotatebox{-90}{\FigureFile(60mm,60mm){fig1_color.eps}}
\end{center}
\caption{The 2-10 keV XIS\,3 lightcurve of MCG -5-23-16, shown as black 
circles. The observation started on 7 December 2005, with a total 
exposure time (after screening) of $\sim100$~ks. The Suzaku lightcurve 
has been binned into orbital length bins of 5760s. Overlaid in red is the 
XMM-Newton EPIC-pn lightcurve, which is simultaneous for part of the 
observation.}
\end{figure}

\section{The Suzaku Observation of MCG-5-23-16}

MCG\,-5-23-16 was observed by Suzaku between 7-10 December 2001 with a total 
duration of 220\,ks (see Table\,1 for a summary of observations). Events files 
from version 0.7
of the Suzaku pipeline processing were used.
\footnote{Version 0 processing is a simplified processing applied to the Suzaku  
data obtained during the SWG phase, for the purpose of establishing the
detector calibration as quick as possible. Some processes that are not
critical for most of the initial scientific studies, e.g., aspect
correction, fine tuning of the event time tagging of the XIS data, are
skipped in version 0 processing, and hence, the quality of the products
is limited in these directions, compared with the official data
supplied to the guest observers. As of 2006 July, version 0.7 is the
latest, where the absolute energy scale of $\sim 5$~eV is achieved for
the XIS data.}
All events files were further screened within \textsc{xselect} to 
exclude data within the SAA (South Atlantic Anomaly) 
as well as excluding data with an Earth elevation angle 
(ELV) $<5$ degrees. Data with Earth day-time elevation angles 
(DYE\_ELV) less than 20 degrees were also excluded. 
Furthermore data within 256s of the SAA were excluded 
from the XIS and within 500s of the SAA for the HXD (using the T\_SAA\_HXD parameter 
within the house-keeping files). Cut-off rigidity (COR) 
criteria of $>8$\,GeV/c for the HXD data and $>6$\,GeV/c for the XIS were used. 

\begin{table*}
 \caption{Summary of Observations}
  \begin{center}
    \begin{tabular}{lccc}
\hline
     Instrument & TSTART (UT) & TSTOP (UT) & Exposure (ks) \\
\hline   
     Suzaku/XIS & 2005/12/07 22:59:49 & 2005/12/10 11:48:54 & 98.1 \\
     Suzaku/PIN & 2005/12/07 22:40:00 & 2005/12/10 11:50:50 & 71.4 \\
     Suzaku/GSO & 2005/12/07 22:40:00 & 2005/12/10 11:50:50 & 18.3 \\
     XMM-Newton/PN & 2005/12/08 21:11:33 & 2005/12/10 06:57:51 & 96.2 \\
     Chandra/HETG  & 2005/12/08 17:41:30 & 2005/12/09 02:33:59 \\
     Chandra/HETG  & 2005/12/09 20:52:11 & 2005/12/10 03:00:10 & 49.5 \\
     RXTE/PCA      & 2005/12/09 13:16:00 & 2005/12/09 20:30:00 & 12.8 \\
     Swift/XRT     & 2005/12/09 00:03:00 & 2005/12/09 09:54:57 & 10.1 \\
\hline
     \end{tabular}
  \end{center}
\end{table*}

\subsection{XIS Data Analysis}

For the XIS, only good events with grades 0,2,3,4 and 6 were used, 
while hot and flickering pixels were removed from the XIS images using the 
\textsc{cleansis} script. Time intervals effected by telemetry saturation were 
also removed.  
The XIS was set to normal clocking mode and the data format
was either in the $3\times3$ or $5\times5$ edit modes (corresponding to 
9 and 25 pixel pulse height arrays respectively that are telemetered 
to the ground). 
The XIS pulse height data for  
each X-ray event were converted to
PI (Pulse Invariant) channels using the \textsc{xispi} software  
version 2005/12/26 and CTI parameters were used 
from 2006/05/22. 

The 4 XIS source spectra (see \citet{Koyama06} for details of the XIS) 
were extracted from circular source regions of 4.34\arcm\ (6mm) 
radius centered on the source, which was observed 
off-axis in the HXD nominal pointing position. Background spectra were 
extracted from three 2.5\arcm\ circles offset from the source region.  
The source and background extraction regions were 
also chosen to avoid the calibration sources on the corners of the CCD chips. XIS 
response files (rmfs) provided by the instrument teams were used (dated 
2006/02/13), while ancillary 
response files (arfs) appropriate for the HXD nominal pointing 
were chosen dated as of 2006/04/15,  
matching the size (6mm) of the source extraction regions 
used. For the front illuminated XIS chips (XIS 0,2,3) 
data over the energy range 0.6--10\,keV were used, while for the softer 
back-illuminated XIS\,1 chip data from 0.4--8\,keV were used. Data around the CCD 
Si K edge from 1.7--1.95\,keV were ignored in all 4 XIS chips, due to 
uncertain calibration at this energy at the time of writing. A total source exposure 
(after screening) of 98.1 \,ks was obtained for the 4 XIS chips.  The 2-10\,keV source 
lightcurve for the XIS\,3 is shown in Figure\,1.

The XIS spectra were corrected 
for the hydrocarbon (C$_{24}$H$_{38}$O$_{4}$) 
contamination on the optical blocking filter, by including 
an extra absorption column due to Carbon and Oxygen in all the spectral fits. 
The column densities for each detector were calculated 
based on the date of the observation and the off-axis position of the source. 
The Carbon column densities ($N_{\rm C}$) used were $1.64\times10^{18}$, $2.18\times10^{18}$, 
$2.82\times10^{18}$ and $4.40\times10^{18}$\,atoms\,cm$^{-2}$ for XIS 0,1,2,3 
respectively, with the ratio of C/O column densities set to 6. 
The additional soft X-ray absorption due to the 
hydrocarbon contamination was included as a fixed spectral 
component using the \textsc{varabs} absorption model in all the spectral fits. 

The net source count rates obtained for the 4 XIS detectors are 
$2.843\pm0.006$\,counts\,s$^{-1}$, $3.081\pm0.006$\,counts\,s$^{-1}$, 
$3.403\pm0.007$\,counts\,s$^{-1}$ and $3.316\pm0.007$\,counts\,s$^{-1}$ for XIS\,0--3 
respectively, with background typically 1\% of the source rate.  
Given the large number of source counts available in the observation, the XIS 
source spectra were binned to a minimum of 200 counts per bin to enable 
the use of the $\chi^{2}$ minimization technique. 

\subsection{HXD Data Analysis}

\subsubsection{HXD/PIN}

The source spectrum was extracted from the cleaned HXD/PIN 
events files, processed with the screening criteria described above. 
The HXD/PIN instrumental background spectrum was generated from a time dependent 
model provided by the HXD instrument team. The model
utilized the count rate of upper discriminators as the measure
of cosmic-ray flux that passes through the silicon PIN diode and
provide background spectra based upon a database of
non-X-ray background observations made by the PIN diode to date
\citep{Kokubun06}. 
Both the source and background spectra were made with identical GTIs 
(good time intervals) and the source exposure was corrected 
for detector deadtime (which is $<5$\%). A detailed description of the 
PIN detector deadtime is given in \citet{Kokubun06}. 
The net exposure time of the 
PIN source spectrum was 71.4\,ks after deadtime correction. Note that the 
background spectral model was generated with
$10\times$ the actual background count rate in order to minimize the photon 
noise on the background; this has been accounted for 
by increasing the effective exposure time of the background spectra by a factor 
of $\times10$.  
The HXD/PIN response file dated 
2006/08/14 for the HXD nominal position was used for these spectral fits. 
Instrumental and performance 
details of the HXD are discussed in \citet{Takahashi06} and 
\citet{Kokubun06}. 

In addition a spectrum of the 
cosmic X-ray background (CXB) \citep{Boldt87, Gruber99} 
was also simulated with the HXD/PIN. The form of the CXB was taken as 
$9.0\times10^{-9}\times({\rm E}/3\,\rm{keV})^{-0.29}\times e^{(-\rm{E}/40\,\rm{keV})}$\,erg\,cm$^{-2}$\,s$^{-1}$\,sr$^{-1}$\,keV$^{-1}$. 
When normalized to the field of view of the 
HXD/PIN instrument 
the effective flux of the CXB component is 
$9.0\times10^{-12}$\,erg\,cm$^{-2}$\,s$^{-1}$ in the 12--50 keV band.  
Note the flux of MCG\,-5-23-16 measured by the HXD over the same 
band is $1.6\times10^{-10}$\,erg\,cm$^{-2}$\,s$^{-1}$, i.e. the CXB component 
represents only 6\% of the net source flux measured by the HXD/PIN. Note that 
there is some uncertainty in the absolute flux level of the CXB component measured 
between missions, 
for instance \citet{Chur06} find the CXB normalization from {\it Integral} to be about 10\% 
higher than measured by \citet{Gruber99} from the HEAO-1 data. However this level of 
uncertainty is much smaller than the net source flux of MCG\,-5-23-16, at the level of $<1\%$. 
Furthermore the spatial fluctuation of the CXB also has a negligible effect
($<1$\% of the net source flux), for instance 
\citet{Kushino02} measure a $6.5\pm0.6$\% fluctuation based on the 2--10\,keV 
ASCA GIS data, which has a total field of view similar to Suzaku HXD/PIN. 

\begin{figure}
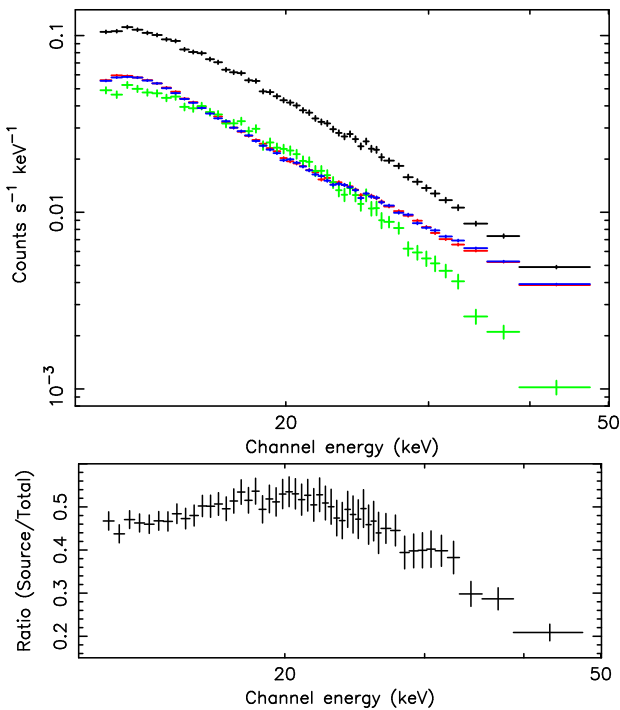

\begin{center}
\rotatebox{-90}{\FigureFile(60mm,70mm){fig2a_color.eps}}
\rotatebox{-90}{\FigureFile(33mm,50mm){fig2b.eps}}
\end{center}
\caption{HXD/PIN spectra for the Suzaku observation of MCG\,-5-23-16. The upper panel 
shows the total spectrum (source plus background, black). 
The points in red and blue show two independent 
background models for HXD/PIN, the red points are the ones adopted in the paper. 
The two backgrounds show very good agreement. 
The net source counts (i.e. total minus model background) are shown in green. Note the 
source is detected in each data bin at a minimum of $10\sigma$ above the background. 
The lower panel shows the ratio of the net source spectrum to the total 
spectrum, which is $>20$\% over the 12--50 keV range.} 
\end{figure}

\begin{figure}
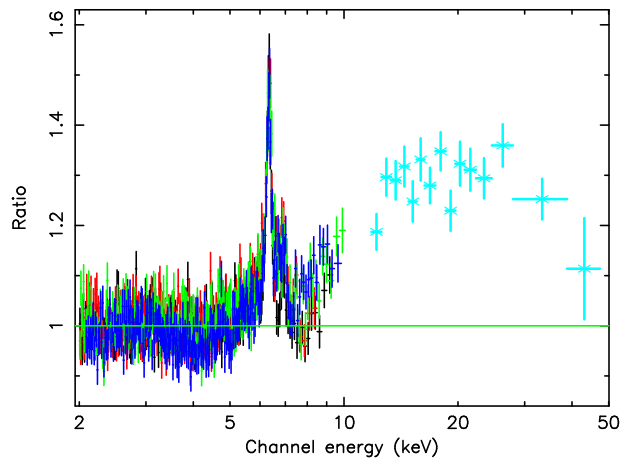

\begin{center}
\rotatebox{-90}{\FigureFile(60mm,60mm){fig3_color.eps}}
\end{center}
\caption{The four Suzaku XIS spectra (XIS\,0, black; XIS\,1, red; XIS\,2; green; 
XIS\,3; blue) and HXD/PIN (light blue stars) over the 2-50\,keV range plotted as 
a ratio to an absorbed power-law of photon index $\Gamma=1.8$. 
A 15\% cross normalization factor has been accounted for between HXD/PIN and the XIS 
(the HXD/PIN is thought to be 13--15\% higher with the current calibration). 
The cross-normalization between the 4 XIS detectors 
above 2\,keV agrees within 5\% and the photon index is consistent within 
$\Gamma=\pm0.05$ for the 4 XIS spectra. 
There is an excess of counts above 12 keV in the HXD/PIN compared to the XIS, most 
likely due to a Compton reflection hump.}
\end{figure}

The HXD/PIN spectrum is shown in Figure 2, plotted from 12--50\,keV. 
At present we exclude the HXD/PIN data above 50 keV, as a detailed study 
of the background systematics for HXD/PIN is on-going above this energy. 
Figure 2 shows the total (source + observed background) spectrum and 
two independent background 
model spectra provided by the HXD instrument team 
(the red and blue points), both of which include the instrumental (non X-ray)
background and the contribution from the CXB. The net source spectrum is 
plotted (green 
points), which shows the total spectrum minus the background model spectrum (the 
red points are the background model adopted in this paper).  
The agreement between the two independent background models
is excellent, the spectral fits derived for MCG\,-5-23-16 
produces consistent results (within the statistical errors quoted in the paper) 
for each background model. The lower panel of Figure 2 
shows the ratio of net source counts to the total observed PIN counts. 
Thus the net source counts are $>20$\% of the total counts for each bin from 12--50\,keV, 
while the source is detected at $>10\sigma$ above the background in each bin in the spectrum. 
The resulting net PIN source count rate from 12--50\,keV 
was $0.454\pm0.003$\,counts\,s$^{-1}$, compared to the 
PIN background rate of $0.494\pm0.001$\,counts\,s$^{-1}$.

\subsubsection{HXD/GSO}

For the HXD/GSO, background spectra were extracted from a Suzaku observation 
of the 
cluster Abell\,1795 which occurred from December 10--13, 2005, 
just after the MCG\,-5-23-16 observation. Abell 1795 was not detected in the 
HXD/GSO and is therefore suitable for use as a background template 
given the close proximity in time of the two observations. 
Both the GSO background from Abell\,1795 and the MCG\,-5-23-16 source spectra 
were extracted from orbits that contained no SAA passage, which have the most reliable 
(and least variable) GSO background. Matching orbits (i.e. at the same time of day) were 
also chosen between the MCG\,-5-23-16 and Abell\,1795 observations. 
After applying this additional screening, the net exposure times for the GSO source and 
background spectra were then 18.3\,ks and 28.6\,ks respectively.   
The source was detected at the $15\sigma$ statistical level over the GSO background in the 
50-100\,keV band and at a level of 5.5\% over the detector background. As a study 
of the background systematics of the GSO is still on-going, unless otherwise 
stated we do not include the GSO data in the subsequent spectral fits.  
Nonetheless a plot of the GSO detection between 50--100\,keV and of the 
HXD/PIN and XIS spectra is shown in Figure\,4, compared to an absorbed power-law 
spectrum (see description later).  
Above 100\,keV the GSO background systematics are more uncertain and subject to 
further study, so a reliable detection of this source 
above this energy cannot be made at the present time.

\subsection{Simultaneous Observations With Other Observatories}

Simultaneous observations were also conducted with XMM-Newton, 
Chandra (HETG), RXTE and Swift covering part of the Suzaku observation. 
The details of the observations are shown in Table\,1. Figure 1 shows the 
lightcurve from 
Suzaku XIS\,3 compared to the XMM-Newton EPIC-pn observation, with the 100\,ks XMM-Newton 
observation overlapping with part of the Suzaku observation. The \textsc{pattern=0} 
data only were used for the EPIC-pn. 
A more detailed description of the analysis of the simultaneous observations 
will appear in a subsequent paper (Braito et al. 2006, in preparation); the 
purpose of this paper is to describe the Suzaku observation especially the 
properties of the iron K line and the hard X-ray emission and 
reflection component above 10\,keV. Nonetheless we refer below to the 
Chandra HETG data for the measurement of the width of the 
iron K$\alpha$ line core, 
to XMM-Newton to check the consistency of the iron K line profile with Suzaku 
and to RXTE as an independent check of the reflection component. 

\section{Broad-Band X-ray Spectral Analysis}

\subsection{Initial Spectral Fitting}

We first concentrate our analysis on the time-averaged 
MCG\,-5-23-16 spectrum, from the whole Suzaku observation. 
The XIS and HXD/PIN background subtracted spectra were fitted using 
\textsc{xspec v11.3},
including data over the energy range 0.4--50\,keV. 
A Galactic absorption column of $N_{H}=8.0\times10^{20}$~cm$^{-2}$ 
\citep{Dickey90} was included in all the fits  
and spectral parameters are quoted in the rest-frame of MCG\,-5-23-16 at 
$z=0.008486$. All errors are quoted at 90\% confidence for one 
interesting parameter (corresponding to $\Delta\chi^{2}=2.7$) unless otherwise 
stated. A cosmology of $H_{0}=70$\,km\,s$^{-1}$\,Mpc$^{-1}$ and $\Lambda_{0}=0.73$ is 
assumed throughout.

Initially we fitted the 4 XIS spectra and HXD data using a simple absorbed 
power-law model in the 2--5\,keV band in order for the continuum parameters
not to be 
affected by the iron K$\alpha$ line at 6.4\,keV, the reflection component at 
high energies, as well as any soft X-ray spectral complexity below 1\,keV. 
The continuum was fitted with a photon index $\Gamma\sim1.8$ and an 
absorption column of $1.6\times10^{22}$\,cm$^{-2}$. 
The relative normalization of the HXD/PIN was initially fixed at 15\% above that of the 
XIS, as has been found to date from calibration observations 
of the Crab (Kokubun et al. 2006). Figure 3 shows the 
spectrum and data/model ratio of the 4 XIS spectra as well as the HXD/PIN data 
between 2--50\,keV against the absorbed power-law continuum. 
The photon indices of the 4 XIS spectra are all in good agreement to within 
$\pm0.05$ of each other; 
these are $\Gamma=1.84\pm0.03$ (for XIS\,0), $\Gamma=1.79\pm0.04$ (XIS\,1), 
$\Gamma=1.82\pm0.03$ (XIS\,2) and $\Gamma=1.82\pm0.03$ (XIS\,3). 
The cross normalization between all 4 XIS detectors is also in good 
agreement to $\pm5$\%. An excess of flux in the HXD/PIN compared to the XIS spectra 
is clearly apparent. 

\begin{figure}
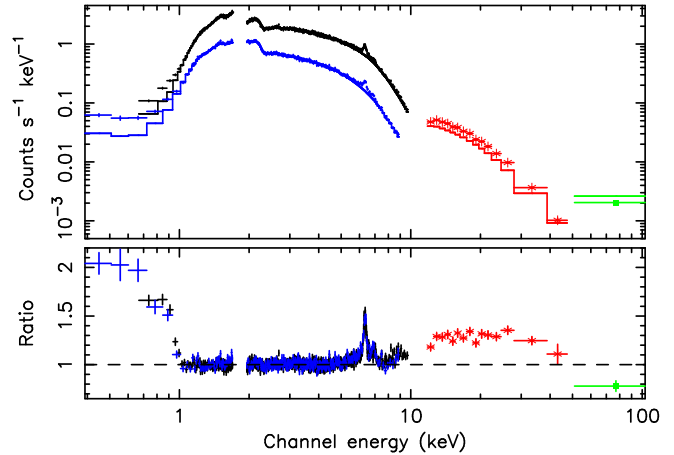

\begin{center}
\rotatebox{-90}{\FigureFile(60mm,60mm){fig4_color.eps}}
\end{center}
\caption{The broad-band (0.4-100\,keV) Suzaku spectrum of MCG -5-23-16. 
The upper panel shows the data, plotted against an absorbed power-law 
model of photon index $\Gamma=1.8$ (solid line) and column density 
$1.6\times10^{22}$\,cm$^{-2}$, fitted over the 2-5\,keV band.
The lower panel shows the data/model ratio 
residuals to this power-law fit. The data from XIS\,FI chips are 
shown in black, the XIS 1 in blue, the HXD/PIN as red stars 
and the HXD/GSO as green squares. 
Clear deviations in the iron K-shell band are 
apparent between 6-7 keV, while an excess of counts (due to a soft excess) is 
present below 1 keV. A hard excess is seen in the HXD spectrum above 10 keV, 
due to the presence of a reflection component in this object.}
\end{figure}

Due to the good agreement between 
the XIS spectra, data from the three front illuminated XIS\,0, XIS\,2 
and XIS\,3 chips were co-added (hereafter referred to as XIS\,FI) 
in order to maximize signal to noise.
We do not combine the XIS\,1 which is a back 
illuminated chip and its response 
is better suited to the soft X-ray spectrum, 
with lower effective area at high energies. 
Figure 4 shows the spectrum and data/model residuals to the 
entire Suzaku spectrum from XIS\,FI, XIS\,1, 
HXD/PIN and HXD/GSO over the 0.4-100\,keV 
range against the simple absorbed power-law model. Note 
we include the GSO datapoint here for comparison, but it is 
not used in the spectral fitting. An excess of counts above the 
power-law continuum is seen above 10\,keV likely due to Compton 
reflection in the X-ray spectrum, while the spectrum appears to turn over at the 
very highest energies, perhaps due to a high energy exponential cut-off. 
At lower energies a soft excess is seen below 1\,keV 
(especially in the softer XIS\,1 instrument), which may be due to an unabsorbed 
component in scattered emission associated with this Seyfert 1.9 galaxy.

\subsection{The Baseline Continuum Model}

We then constructed a baseline model which can fit the spectrum of MCG\,-5-23-16. 
Initially we concentrated on the joint fit between 
XIS\,FI and HXD/PIN from 0.4--50\,keV. For the continuum, we modeled the spectrum 
with an absorbed power-law component including an exponential cut-off at high 
energies of the form, $N_{\rm PL} E^{-\Gamma} \times e^{-E/E_{\rm cut}}$; 
where $N_{\rm PL}$ is the normalization of the power-law (in photons\,cm$^{-2}$\,s$^{-1}$\,keV$^{-1}$ 
at 1 keV), $E$ is the energy (in keV), $\Gamma$ is the photon index and $E_{\rm cut}$ 
is the e-folding cut-off energy (in keV). 

This primary power-law continuum is absorbed by two layers of neutral gas; an 
absorbing layer at the redshift of MCG\,-5-23-16 and the Galactic absorption column of 
$8\times10^{20}$\,cm$^{-2}$ \citep{Dickey90}. A second soft X-ray 
power-law component was included, absorbed by only the Galactic column, in order 
to model the soft excess. 
Abundances were set those of \citet{Wilms00} using the cross-sections of 
\citet{BM92}. 
Furthermore a reflection 
component produced by Compton down-scattering of X-rays off neutral material
was included, using the \textsc{pexrav} model within \textsc{xspec} \citep{MZ95}. 
The reflection component is absorbed by the Galactic column only. 
The continuum parameters ($\Gamma$, $N_{\rm PL}$ and $E_{\rm cut}$) were 
tied to those of the incident cut-off power-law. The inclination ($\theta$) 
of the reflector was set to 45\degg, while the Fe abundance ($A_{\rm Fe}$) 
was allowed to vary. The solid angle ($\Omega$) subtended by the Compton reflector was 
determined by the parameter $R=\Omega/2\pi$. The best-fit parameters are 
shown in Table\,2. The primary power-law has a photon index of $\Gamma=1.95\pm0.03$ after 
the inclusion of the reflection component, 
with a high energy cut-off of $E_{\rm cut}>170$\,keV, absorbed by 
a neutral column density of $N_{\rm H}=1.65\pm0.03\times10^{22}$\,cm$^{-2}$. 
The soft, unabsorbed power-law required to fit the soft excess has a steeper photon index of 
$\Gamma=3.1\pm0.3$. The properties of the iron K line and the reflection component 
are described below. 

\begin{table*}
 \caption{Spectral Fit Parameters. Note $^{a}$ normalization of power-law in units 
$10^{-2}$\,photons\,cm$^{-2}$\,s$^{-1}$\,keV$^{-1}$ at 1 keV; $^{b}$ X-ray flux in units of 
$10^{-11}$\,erg\,cm$^{-2}$\,s$^{-1}$; $^{c}$ column density in units of $10^{22}$\,cm$^{-2}$; 
$^{d}$ normalization (flux) of iron line in units $10^{-5}$\, photons\,cm$^{-2}$\,s$^{-1}$; 
$^{e}$ inner disk radius in gravitational radii. $^{f}$ donates parameter is fixed in fit.}
\label{tab:second}
  \begin{center}
    \begin{tabular}{llccc}
\hline
     & Parameter & Mean & High & Low \\
\hline   
     Continuum  & $\Gamma$ & $1.95\pm0.03$ & $1.92\pm0.03$ & $1.93\pm0.03$ \\
                & $N_{\rm PL}$$^{a}$ & $3.2\pm0.1$ & $3.5\pm0.1$ & $2.6\pm0.1$ \\
                & $E_{\rm cut}$ (keV) & $>170$ & $200^{f}$ & $200^{f}$ \\
                & ${\rm Flux}_{2-10}$$^{b}$ & 8.76 & 9.91 & 7.12 \\
                & ${\rm Flux}_{15-100}$$^{b}$ & 19.0 & 20.0 & 14.1 \\
                & $N_{\rm H}$$^{c}$ & $1.65\pm0.03$ \\
     Scattered  & $\Gamma$ & $3.1\pm0.3$ \\
                & $N_{\rm PL}$$^{a}$ & $1.5\pm0.2\times10^{-2}$ \\
     Reflection & $R$ & $1.1\pm0.2$ & $1.2\pm0.2$ & $1.3\pm0.2$ \\
                & Abund (Fe) & $0.40\pm0.13$ & $0.5^{f}$ & $0.5^{f}$ \\
     Narrow     & E (keV) & $6.400\pm0.007$ & $6.40\pm0.01$ & $6.40\pm0.01$ \\
     Gaussian   & $\sigma$ (eV) & $<48$ & $10^{f}$ & $10^{f}$ \\
                & $N_{\rm line}$$^{d}$ & $6.5\pm0.5$ & $6.5\pm2.0$ & $5.4^{+1.7}_{-2.6}$ \\
                & EW (eV) & $70\pm6$ & $65\pm19$ & $91^{+28}_{-44}$ \\
     Broad      & E (keV) & $6.39\pm0.10$ & $6.4\pm0.1$ & $6.4\pm0.1$ \\
     Gaussian   & $\sigma$ (keV) & $0.44\pm0.12$ & $0.4^{f}$ & $0.4^{f}$ \\
                & $N_{\rm line}$$^{d}$ & $6.1\pm1.6$ & $5.6^{+3.1}_{-3.6}$ & $6.8^{+3.4}_{-2.9}$ \\
                & EW (eV) & $62\pm17$ & 53 & 92 \\
     or Diskline & E (keV) & 6.40$^{f}$ \\
    (XMM+Suzaku) & $R_{\rm in}$$^{e}$ & $37^{+25}_{-10}$ \\
                 & $\theta$ & $53^{+7}_{-9}$ \\
                 & $N_{\rm line}$$^{d}$ & $5.1\pm0.9$ \\
                 & EW (eV) & $60\pm11$ \\
\hline
     \end{tabular}
  \end{center}
\end{table*}

\subsection{The Iron K Line Profile}

The data/model residuals of the XIS\,FI spectrum in the iron K band to this 
continuum model are shown in Figure 5 (without the inclusion of the reflection 
component). For comparison, the 
XMM-Newton EPIC-pn data are also plotted, 
which shows that the iron K line profiles obtained 
by XMM-Newton and Suzaku are in excellent agreement. The data from both missions show 
a strong iron K$\alpha$ core at 6.4\,keV, an excess of counts both red-wards and 
blue-wards of 6.4 keV, a weak peak at 7\,keV due to the iron K$\beta$ line and 
a drop at 7.1\,keV, likely due to the neutral Fe K edge associated with the 
Compton reflector. 

\begin{figure}
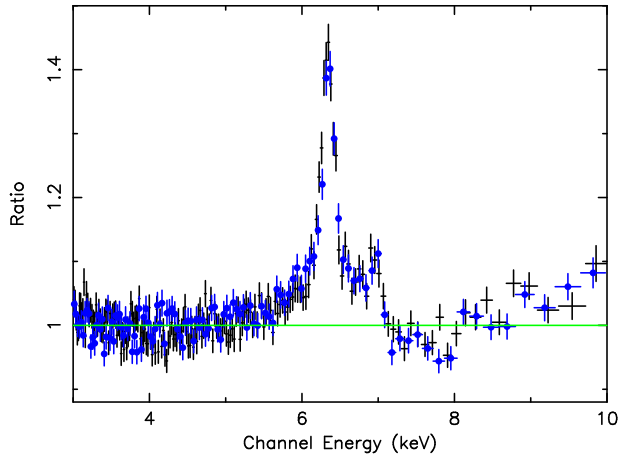

\begin{center}
\rotatebox{-90}{\FigureFile(60mm,60mm){fig5_color.eps}}
\end{center}
\caption{The iron line profile of MCG -5-23-16, plotted as a ratio against a 
power-law of photon index $\Gamma=1.8$. The data from Suzaku XIS\,FI is shown 
in black and for comparison the blue circles show the data from the XMM-Newton 
EPIC-pn observation. 
Both observations show a narrow iron K$\alpha$ core at 6.4 keV, a red-wing below 
6.4 keV, a peak at 7.05\,keV due to Fe K$\beta$. The edge at 7.1 keV is due to the 
reflection component.}
\end{figure}

The iron K line profile was then fitted in several steps. Firstly Compton 
reflection was included, as described above. The resultant line profile to the 
Suzaku data can be seen in Figure 6, panel (a). Then the narrow core to the K$\alpha$ 
line was added to the model, while a narrow K$\beta$ line was also included, 
the line energy of the K$\beta$ line was fixed at 7.056\,keV, 
with the line flux set equal to 12\% of the K$\alpha$ as expected for 
neutral iron. For completeness, we also add a small Compton shoulder to the 
iron K$\alpha$ line, represented by a narrow Gaussian centered at 6.3\,keV, with 
normalization fixed to 20\% of the K$\alpha$ flux, expected if the emission originates 
from Compton-thick material \citep{Matt02}. 
After the inclusion of the narrow lines, a clear 
excess of counts in the Suzaku data can be seen around 6.4\,keV 
(Figure 6, panel b). This excess can modeled by either a broad 
Gaussian line or a diskline profile from an accretion disk \citep{Fabian89}. 
Figure 6(c) shows the residuals in the iron K band after fitting the 
iron K line with the narrow K$\alpha$ and K$\beta$ lines, as well as a broad 
iron line component. The fit statistic obtained 
is good considering the high statistical quality of the data 
($\chi^{2}/{\rm dof}=1983/1879$ for the disk-line model, 
where dof is the number of degrees of freedom). 



\subsubsection{The Narrow K$\alpha$ Line}
 
The narrow K$\alpha$ core has an equivalent width of $70\pm6$\,eV, centered at 
$6.391\pm0.007$\,keV, with a measured width of $\sigma=51\pm12$\,eV. The $^{55}{\rm Fe}$
calibration source located on the corners of the XIS chips can be used as an accurate calibrator
of the energy and intrinsic width of the iron line. The $^{55}{\rm Fe}$ source produces lines 
from Mn K$\alpha$ (${\rm K}\alpha_{1}$ \& ${\rm K}\alpha_{2}$ 
at 5.899\,keV and 5.888\,keV respectively 
with a branching ratio of 2:1). From measuring the lines in the calibration source, we find 
that the line energy is shifted redwards by $9\pm2$\,eV, while there is some residual 
width (after the instrumental response function has been accounted for) 
in the calibration lines of $\sigma=41\pm2$\,eV. 
Therefore the intrinsic width ($\sigma_{\rm int}$) of the iron K$\alpha$ line is simply 
$\sigma_{\rm int}^{2}=\sigma_{\rm meas}^{2} - \sigma_{\rm cal}^{2}$ (where 
$\sigma_{\rm meas}$ is the measured width and $\sigma_{\rm cal}$ is the residual width of the 
calibration lines). 

The intrinsic width is thus formally consistent with being unresolved, while a statistical 
upper limit can be placed on the width of $\sigma_{\rm int}<48$\,eV, 
corresponding to a velocity width of $\sigma_{\rm v}<2200$\,km\,s$^{-1}$. 
The corrected energy of the narrow iron line 
is then $6.400\pm0.007$\,keV, consistent with a superposition of the 
neutral K$\alpha_{1}$ and K$\alpha_{2}$ iron lines at 6.404\,keV and 6.391\,keV 
respectively. In comparison, the line width and energy obtained from the simultaneous 
Chandra/HETG observation are $\sigma<43$\,eV 
($\sigma_{\rm v}<2000$\,km\,s$^{-1}$) and $6.405\pm0.015$\,keV (Braito et al. 2006)
which is in excellent agreement. Note that the residual broadening in the XIS 
response function has a negligible effect on the broad line parameters 
determined below.

\begin{figure}
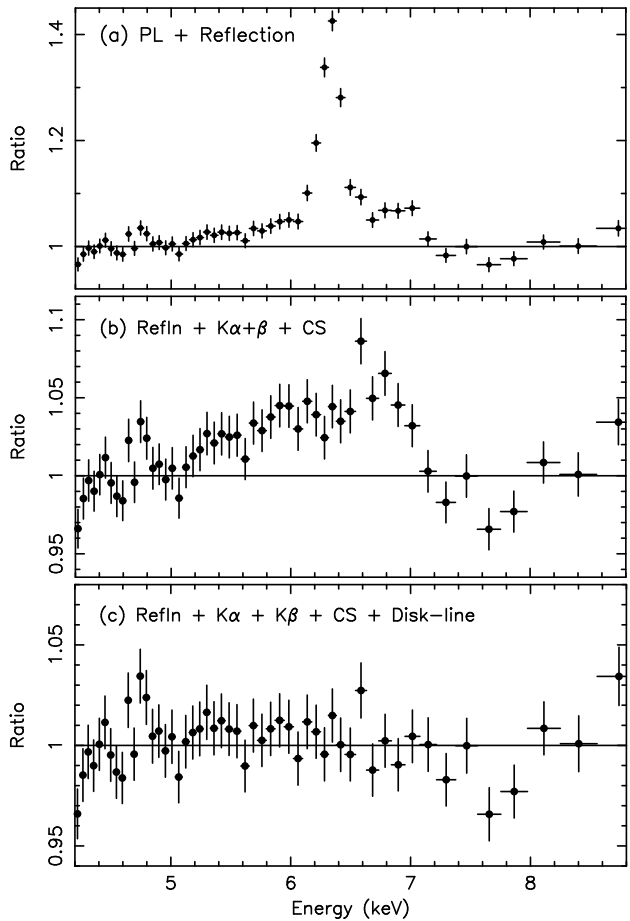

\begin{center}
\rotatebox{-90}{\FigureFile(38mm,110mm){fig6a.eps}}
\rotatebox{-90}{\FigureFile(38mm,110mm){fig6b.eps}}
\rotatebox{-90}{\FigureFile(45mm,110mm){fig6c.eps}}
\end{center}
\caption{Data/model ratio residuals of the spectrum of 
MCG\,-5-23-16, in the iron K-shell band. 
Panel (a) shows the residuals after fitting a 
power-law with a  reflection component of $R=1.1$. 
Panel (b) shows the residuals after the narrow K$\alpha$ and 
K$\beta$ Fe emission lines at 6.40 keV and 7.05 keV have been fitted. A broad 
excess is observed both red-wards and blue-wards of 6.4 keV.  
Panel (c) shows the residuals after the addition of a broad iron K$\alpha$ 
line centered at 6.4 keV. The fit in the iron K band is now good, for 
completeness a weak Compton shoulder to the narrow 
Fe K$\alpha$ line core has also been included. Note that the spectra have been 
re-binned by a factor of $\times25$ for clarity.}
\end{figure}

\subsubsection{The Broad Iron Line}

When modeled by a Gaussian, the broad component of the line is also centered near 6.4\,keV 
($E=6.43\pm0.10$\,keV), with an equivalent width of $62\pm17$\,eV and an intrinsic 
width of $\sigma=440\pm120$\,eV or $\sigma_{\rm v}\sim20000$\,km\,s$^{-1}$. 
This large velocity indicates that the broad line must originate from close 
to the black hole in MCG\,-5-23-16. Therefore the broad iron line was 
modeled with an emission profile from an accretion disk around a Schwarzschild 
black hole (\textsc{diskline}, \citet{Fabian89}). The outer radius of 
the diskline was fixed to $400R_{\rm g}$ (where $R_{\rm g}=GM/c^{2}$), while 
it was assumed that the iron K emission originates from neutral matter at 6.4\,keV. 
The disk emissivity is parameterized in the form $r^{-q}$, where 
$q$ is the emissivity index and $r$ is the disk radius. Initially 
the emissivity was fixed at $q=3$. 
The disk inclination angle is then constrained to be $50^{+32}_{-10}$\degg, 
the inner radius of the disk is $26^{+35}_{-8}{\rm R}_{\rm g}$, 
while the line EW is $63\pm16$\,eV. If the constraint on the 
disk emissivity is relaxed, then a disk inner radius of $6R_{\rm g}$ is 
allowed, with a flatter emissivity of $q=2.0^{+0.4}_{-0.7}$. 
Overall the fit statistic is good, with $\chi^{2}/{\rm dof}=1983/1879$ (Table 3, model 1); 
if the broad line is not included and the data are refitted then the 
fit is considerably worse ($\chi^{2}/{\rm dof}=2027/1882$, Table 3, model 2). Thus the broad 
line is required in the data at $>99.99$\% confidence. We also note that a simple 
broad Gaussian profile produces a very slightly worse fit than a diskline profile
($\chi^{2}/{\rm dof}=1987/1879$). 
In order to obtain tighter constraints on the diskline, a combined fit was then performed 
with the Suzaku and XMM-Newton EPIC-pn data. For this fit the inner radius 
of the disk is $37^{+25}_{-10}{\rm R}_{\rm g}$ 
(for a fixed emissivity of $q=3$), the inclination angle is 
$53^{+7}_{-9}$\degg, while the line EW is $60\pm11$\,eV. A summary of the 
line parameters is shown in Table\,2.

Finally it has been previously suggested from XMM-Newton studies of the iron line 
profile in AGN that complex absorption (either ionized or partially covering the 
X-ray source) could provide an alternative explanation for the relativistic iron line 
\citep{R04, Pounds04, Turner05}. In order to test this hypothesis we fitted an 
iron K-shell 
bound--free absorption edge constrained between 7.1--9.3\,keV (the known energy 
range from Fe\,\textsc{i} up to Fe\,\textsc{xxvi}), in order to mimic the additional 
opacity of an ionized absorber in the iron K bandpass. While it was
possible to fit a weak iron K-shell edge to the Suzaku data (at $7.2\pm0.1$\,keV with 
an optical depth of $\tau<0.05$), the fit was considerably worse than for a diskline 
($\chi^{2}/{\rm dof}=2020/1880$ with an edge vs. $\chi^{2}/{\rm dof}=1983/1879$ with 
a disk-line). Similarly 
we also attempted to fit the spectrum with a neutral partial covering absorber 
(the \textsc{pcfabs} model in \textsc{xspec}), which also resulted in a 
worse fit ($\chi^{2}/{\rm dof}=2020/1880$) with an absorption column of 
$\sim5\times10^{22}$\,cm$^{-2}$ and a covering fraction of $\sim10$\%. Note 
that there is no evidence for ionized absorption lines and/or edges from lighter 
elements below the iron K band, either from the Suzaku XIS spectrum, or from the 
XMM-Newton EPIC-pm and Chandra/HETG spectra (Braito et al. 2006). Thus other than 
the neutral line of sight column density towards MCG\,-5-23-16 (of 
$1.6\times10^{22}$\,cm$^{-2}$) which has been included in the spectral model, 
there is no evidence of 
complex absorption that effects the detection of the broad iron K line. 

\subsection{Properties of the Reflection Component}

The neutral reflection component is also very well constrained in the above model, 
with a best-fit value of $R=1.1\pm0.2$ with a lower-limit to the cut-off energy of 
$E_{\rm cut}>170$\,keV. Note that if the GSO data-point between 50--100\,keV 
is included then $E_{\rm cut}=270^{+150}_{-70}$\,keV (the reflection 
parameters are unchanged), which is consistent with the Beppo-SAX measurement 
\citep{Perola02, Balestra04} for this source. 
Note the continuum photon index, cutoff energy, iron abundance, and a HXD/XIS constant 
normalization factor are allowed to vary independently along with the 
strength of the reflection hump ($R$). Interestingly the iron 
abundance of the reflector can also be constrained from the ratio of the neutral iron 
K edge to the strength of the Compton hump. The iron abundance was found to be 
sub-Solar, with $A_{\rm Fe}=0.40\pm0.12$ 
relative to the \citet{Wilms00} value.  
We can rule out a solar-abundance of 
iron with a high degree of confidence; fixing the abundance to $A_{\rm Fe}=1$ 
(but still allowing $R$ to vary) worsened the fit statistic by 
$\Delta\chi^{2}=28$, as the Fe K edge in the reflection model at 7.1\,keV is 
then too deep to model the data well. Figure 7 (top panel) 
shows the confidence contours of $R$ against the iron abundance. 

\subsubsection{Consistency Checks}

In the reflection fit, the constant cross normalization factor of the HXD over the
XIS\,FI spectrum was also allowed to vary while fitting the reflection component. 
Figure 7 (lower panel) shows the confidence contours between the constant and 
the reflection fraction (R). We find that in the best-fit described above, there is a small 
constant offset of $1.12\pm0.05$ between the HXD and XIS. This is  
consistent with calibration observations, the XIS+PIN spectrum of the 
Crab \citep{Kokubun06} also shows a factor of $1.15\pm0.01$ offset for a HXD 
nominal pointing and $1.13\pm0.01$ for XIS nominal (with a typical systematic 
uncertainty of 0.02). 
As one would expect, if the offset between the HXD and 
XIS increases, then the amount of reflection required decreases. 
Nonetheless the contours show that even upon allowing for some uncertainty in the 
constant offset between HXD and XIS, 
the parameters of the reflection component are well constrained. Note that the reflection 
parameters (and errors) quoted throughout this paper account for the uncertainty between 
$R$ and the constant offset, as is shown in Figure 7.

\begin{figure}
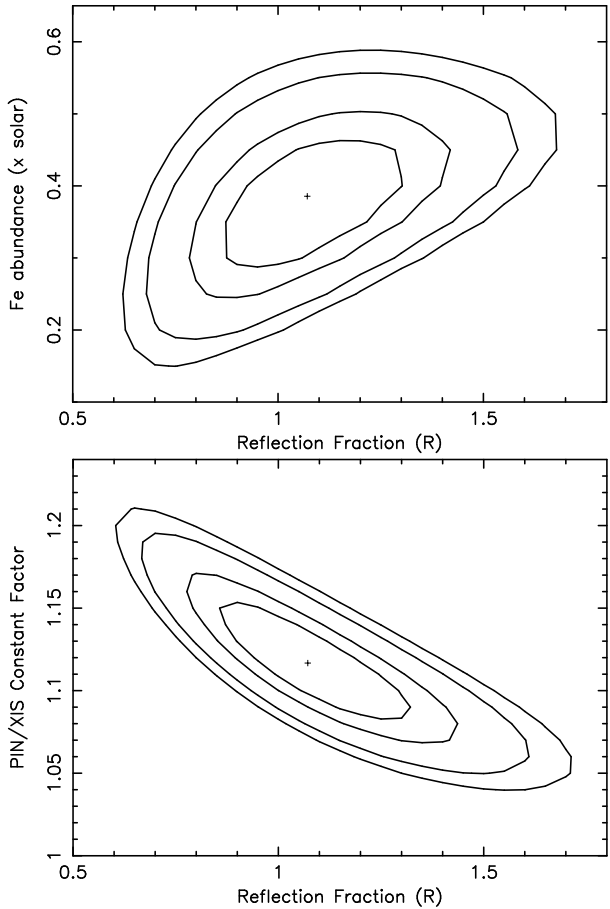

\begin{center}
\rotatebox{-90}{\FigureFile(60mm,110mm){fig7a.eps}}
\rotatebox{-90}{\FigureFile(60mm,110mm){fig7b.eps}}
\end{center}
\caption{Contour plots showing the 68\%, 90\%, 99\% and 99.9\% 
confidence levels of the reflection fraction (R). In the top panel 
we show R versus the iron abundance (compared to the abundances in 
\cite{Wilms00}), demonstrating that the iron abundance is 
sub-Solar. The lower panel plots R versus the constant normalization factor 
between the HXD/PIN and XIS\,FI, which shows that the constant factor is close 
to 1.1 and does not strongly effect the strength of the reflection component.}
\end{figure}

In order to investigate the possible systematics of the HXD background on the reflection parameters, 
the same fits were also performed on the Suzaku data, using instead 
a second independent background model for the HXD/PIN, as was shown in Figure 2 (the 
blue data points). A reflection fraction of $R=1.3\pm0.3$ 
was found to be consistent with the above value within the statistical errors, 
with the other spectral parameters unchanged.  
We also used the simultaneous RXTE/PCA 
spectrum as a mission-independent check on the strength of the reflection component. 
The RXTE PCU\,2 data were reduced with standard extraction criteria, identical 
to those used in Markowitz \& Edelson (2004). 
In fitting the RXTE spectrum, the cut-off energy was fixed at 200\,keV and the iron 
abundance at $0.6\times$~solar, as neither of these parameters can be constrained in the 
PCA data. The iron line parameters were fixed at the Suzaku XIS values, while the 
normalization of RXTE with respect to Suzaku is allowed to vary. With RXTE the 
reflection fraction was constrained to $R=1.0^{+0.7}_{-0.4}$, while $\Gamma=2.05\pm0.12$, 
consistent with the Suzaku observation.  


Overall in the HXD band, the model extrapolated flux from 15-100\,keV of 
MCG\,-5-23-16 is $1.9\times10^{-10}$\,erg\,cm$^{-2}$\,s$^{-1}$. This is consistent 
with the 15-100\,keV flux determined from the Swift BAT survey of 
$1.6\times10^{-10}$\,erg\,cm$^{-2}$\,s$^{-1}$ \citep{Markwardt05} and 
also with a flux of $1.8\times10^{-10}$\,erg\,cm$^{-2}$\,s$^{-1}$ 
obtained from Integral \citep{Bassani06}. The consistency of the flux 
measurements between missions is a further indication of the robustness of the Suzaku HXD 
data.

\subsection{The Soft X-ray Spectrum}

The XIS\,1 were added to the spectral fit to provide better constraints on the 
soft spectrum. The soft unabsorbed power-law component below 1\,keV is found to be 
steep with $\Gamma=3.1\pm0.3$. Furthermore the 
simultaneous XMM-Newton EPIC-pn and MOS data 
also show a soft excess, with a photon index 
of $\Gamma\sim3$. In addition the XIS\,1 data require an emission line at 
$0.92\pm0.01$\,keV, with a flux of 
$1.5\pm0.2\times10^{-5}$\,photons\,cm$^{-2}$\,s$^{-1}$. At this energy the line 
may originate from Ne\,\textsc{ix}, but could also be blended with 
Fe L-shell emission. There are not sufficient counts to determine if there 
is O K-shell emission in the XIS. However the long (100\,ks) XMM-Newton RGS exposure 
simultaneous with the Suzaku observation shows a spectrum dominated by narrow emission
lines below 1\,keV, with lines from N\,\textsc{vii}~Ly\,$\alpha$, 
O\,\textsc{vii} (forbidden), O\,\textsc{viii}~Ly\,$\alpha$, O\,\textsc{vii} RRC,  
Ne\,\textsc{ix} (forbidden) and iron L-shell emission. 
The modeling of XMM-Newton RGS spectrum will be discussed in detail in a subsequent paper 
(Braito et al. 2006). However it seems plausible that the steep soft spectrum is 
the combination of a scattered power-law (with $\Gamma\sim2$ equal to the 
primary continuum) superimposed on 
photoionized emission, similar to other Seyfert 2\,s 
\citep{Turner97, Sambruna01, Kink02, Pounds05, Bianchi05}. Spectral fits 
to the XMM-Newton EPIC and RGS data confirm that this is likely to be the case. 



\section{Variability of the Iron Line and Reflection Component}

\subsection{Short-term Variability}

To test the variability of the iron line and reflection component 
over the timescale of the observation,  
the Suzaku data were split into high and low spectra for both XIS\,FI and 
HXD/PIN respectively. A threshold count rate in XIS\,3 of $>3.7$\,counts\,s$^{-1}$ and 
$<3.1$\,counts\,s$^{-1}$ was used for the high and low flux spectra respectively, 
while spectra were then extracted with identical GTIs for XIS\,0,2, HXD/PIN and the 
HXD/PIN background model. The XIS FI spectra were co-added as above. The 
resultant fluxes for the high and low spectra are then 
$9.9\times10^{-11}$\,erg\,cm$^{-2}$\,s$^{-1}$ and 
$7.1\times10^{-11}$\,erg\,cm$^{-2}$\,s$^{-1}$ respectively in the 2--10\,keV band; 
in the 15-100\,keV band the fluxes are $2.0\times10^{-10}$\,erg\,cm$^{-2}$\,s$^{-1}$
and $1.4\times10^{-10}$\,erg\,cm$^{-2}$\,s$^{-1}$. The data were fitted 
with the same baseline model as above. 

The best fit parameters for the low and high flux spectra are shown in Table\,2. 
In particular the input continuum normalization of the reflection component was 
tied at the mean value for the observation 
($3.2\times10^{-2}$\,photons\,cm$^{-2}$\,s$^{-1}$\,keV$^{-1}$ 
at 1\,keV). Then if the reflection component varies with the continuum, 
an increase in the relative R value (by about 40\%) with respect to 
the mean continuum normalization should be observed in the 
high flux spectrum, compared to the low spectrum.
However the resulting reflection component appears to be constant for the 
high and low state spectra, with values of $R=1.2\pm0.2$ and $R=1.3\pm0.2$ 
respectively. 
Similarly the iron line fluxes of the broad and narrow components 
are consistent with being constant, although some variability cannot be excluded.

As a further check, the difference spectrum of the high minus low flux spectra was 
extracted, using both the XIS and HXD/PIN. The difference spectrum shows the variable 
component of the emission from MCG\,-5-23-16 modified by absorption, with 
the constant components in the spectrum being subtracted. 
The resulting difference spectrum was fitted extremely well with a simple absorbed 
power-law of photon index $\Gamma=1.85\pm0.09$ and 
$N_{\rm H}=1.7\pm0.1\times10^{22}$\,cm$^{-2}$, while the fit statistic is excellent  
($\chi^{2}/{\rm dof}=97.3/156$). Within the errors, this is consistent with the 
continuum parameters derived from the mean spectrum. 
The difference spectrum is plotted in Figure\,8. 
There are no residuals present in the iron K band between 6--7\,keV or any excess counts 
in the HXD/PIN difference spectrum above 10\,keV, consistent with their being no 
variable component to the iron line or the reflection component. 
From the difference spectrum, we can derive an upper-limit to any variable
portion of the reflection spectrum of $R<0.25$, or $<20\%$ of the total reflection
component. The amplitude of the reflection variability is thus smaller than the 
continuum variability. Note  
that there is no excess of counts below 1\,keV in the difference spectrum, also consistent 
with the soft excess being constant. 

\begin{figure}
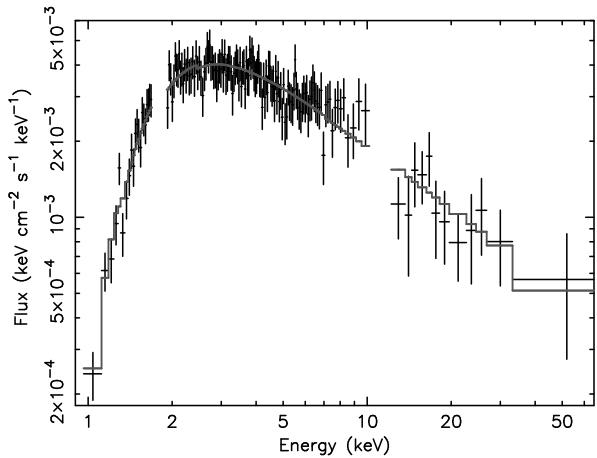

\begin{center}
\rotatebox{-90}{\FigureFile(60mm,110mm){fig8.eps}}
\end{center}
\caption{The difference spectrum between high and low flux states of MCG\,-5-23-16 
in the Suzaku observation. This can be well represented by an absorbed power-law, with 
no iron K emission or reflection. Thus the power-law continuum is variable, while 
there appears to be no variable component of the reflection spectrum.} 
\end{figure}
 
\subsection{Long-Term Variability of the Reflection Component}

The long-term variations in the reflection component was investigated by comparing the 
Suzaku observation with the Beppo-SAX observation in 1998 \citep{Perola02, Balestra04}. 
The value of the reflection fraction in MCG\,-5-23-16 
reported by \citet{Perola02} is $R=0.66^{+0.25}_{-0.20}$, for an inclination 
of $cos\theta=0.9$. When converted to the inclination of 45\degg\ adopted in this 
paper, then their reflection value is $R=0.85^{+0.32}_{-0.25}$, which within the 
errors is consistent with the $R$ value reported here from the Suzaku observation. 
We note that \citet{Balestra04} find a lower $R$ value of 0.45 (for a similar 
inclination of 42\degg) for the same Beppo-SAX observation. Note both of these 
papers assume solar abundances for the reflection component. 

We also performed a combined fit between 
the 1998 Beppo-SAX observation and the 2005 Suzaku observation. 
The continuum parameters (e.g. photon index and cut-off energy) as well 
as the iron line parameters were tied between the Beppo-SAX and the Suzaku datasets, 
allowing only the relative continuum fluxes to vary. The input 
continuum normalization to the reflection component was taken as the average of the 
Suzaku and Beppo-SAX values. 
The cross normalization between SAX/PDS and SAX/MECS was fixed at 0.87. 
The iron abundance with respect to Solar was fixed at 0.6, 
as found earlier. In this 
case the reflection parameters appear consistent with each other, with 
$R=1.0\pm0.2$ and $R=0.7\pm0.2$ for Suzaku and Beppo-SAX respectively. 
The underlying photon index is $\Gamma=1.89\pm0.04$, with a high energy cut-off 
of $200^{+80}_{-50}$\,keV. The flux levels of the observations are 
very similar, the 2-10\,keV fluxes being $8.8\times10^{-11}$\,erg\,cm$^{-2}$\,s$^{-1}$ 
(Suzaku) vs $9.2\times10^{-11}$\,erg\,cm$^{-2}$\,s$^{-1}$ (Beppo-SAX). 
One would expect consistent reflection values, given that the object was 
observed at similar fluxes.

The iron line parameters measured in this paper also appear to be compatible with the 
values measured from previous observations; e.g. see \citet{Dewangan03, Balestra04} 
for comparison. The narrow iron K$\alpha$ line flux is in excellent 
agreement with the previous values 
(between $4-8\times10^{-5}$\,photons\,cm$^{-2}$\,s$^{-1}$), while the EW of the broad 
line measured from Suzaku ($\sim60$\,eV) is also consistent with the 
previous shorter XMM-Newton observations within the errors \citep{Dewangan03, 
Balestra04}. This apparent lack of long term variability 
is perhaps not surprising considering this AGN has remained at a similar flux level 
(typically $7-10\times10^{-11}$\,erg\,cm$^{-2}$\,s$^{-1}$ in the 2-10\,keV band). 
The source was only at a low flux level during the 1980s (see Figure 1, 
\cite{MW04}), as observed during the Ginga 
observations \citep{Nandra94} when the 2-10\,keV flux reached a low of 
$2\times10^{-11}$\,erg\,cm$^{-2}$\,s$^{-1}$. 

\section{Modeling the Reflection Component}

The Suzaku spectrum clearly shows two velocity width 
components to the iron line profile; 
the narrow core being unresolved with Suzaku and even Chandra/HETG, 
while the broad component has a velocity width of $\sim$0.1c. 
Thus it is possible that both components contribute towards the reflection spectrum, 
with the narrow line originating from distant matter (e.g. the putative torus) 
and the broad iron line from 
the accretion disk. Indeed one might expect an accretion disk reflector to contribute 
to the reflection spectrum, given the robust detection of the broad iron line. 

This hypothesis was tested by modeling the 0.4--50\,keV 
Suzaku spectrum with a dual reflection model; 
with an ionized disk reflector (using the \textsc{reflion} model, \cite{RF05}) 
and a neutral distant reflector (using the \textsc{pexrav} model, 
\cite{MZ95}). The narrow iron K$\alpha$, K$\beta$ lines as well as the Compton 
shoulder were added to the model as described previously. 
The broad iron line is included as part of the ionized disk model. 
The ionized disk reflector is blurred by a Kerr metric from around a 
maximally rotating black hole \citep{Laor91}. 
A disk emissivity of $q=3$ is assumed for the disk reflector, with an outer radius 
fixed at $400R_{\rm g}$.  
The best fit parameters for 
the disk reflector are then $R=0.95^{+0.4}_{-0.5}$ for the reflection fraction, 
$\theta=42\pm4$\degg\ for the disk inclination angle with an inner 
radius truncated at $R_{\rm in}=20^{+45}_{-11}\,R_{\rm g}$. 
The underlying continuum is $\Gamma=1.98\pm0.04$, consistent with the previous fits. 
The ionization parameter is consistent with neutral 
or low ionization iron ($\xi<60$\,erg\,cm\,s$^{-1}$). Note that if we allow the 
disk emissivity to vary, then the best fit value is $q=2.3^{+0.9}_{-0.5}$ and
subsequently an inner disk radius extending down to $6R_{\rm g}$ (the last stable orbit 
around a Schwarzschild black hole) or even $1.2R_{\rm g}$ (for a maximal Kerr black hole) 
cannot be ruled out. 

The iron abundance was assumed to be the same value for both the disk and distant 
reflection components; the best fit value found was $A_{\rm Fe}=0.65\pm0.15$.
The neutral distant reflection component has a reflection fraction $R=0.5\pm0.3$ (assuming  
a 45\degg\ inclination), while the equivalent width of the 
narrow iron K$\alpha$ line is $77\pm11$\,eV. For a plane parallel slab, 
illuminated by a continuum of $\Gamma=2$ and viewed at 45\degg\ 
incidence, the predicted iron line equivalent width for a neutral reflector varies between 
60--70\,eV (for $0.5\times$ solar abundance), to 110\,eV (for solar abundance), 
consistent with the measured value for the narrow iron line \citep{Matt02}.  
The overall dual reflection model is shown in Figure 9, while the fit statistic 
of this model fitted to the Suzaku spectrum is good ($\chi^{2}/{\rm dof}=1980/1878$, 
model 3, Table 3.).  
Note that a constant normalization offset factor between HXD/PIN and XIS was 
also included as a fit parameter, this was found to be $1.11\pm0.04$, 
consistent with what was found in the previous section. 

\begin{figure}
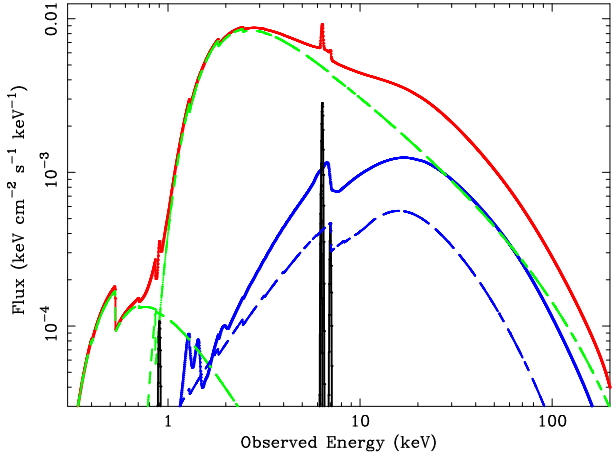

\begin{center}
\rotatebox{-90}{\FigureFile(60mm,60mm){fig9_color.eps}}
\end{center}
\caption{Best fit dual reflection model to the Suzaku spectrum of MCG\,-5-23-16. 
The solid red line shows the total emission from the model, the dashed 
green line shows the primary direct cut-off power-law continuum, while at low 
energies, the dashed line represents the scattered power-law. 
The relativistically blurred ionized disk reflector is plotted with the solid blue 
line, while the distant reflector is shown as a dashed 
blue line. The narrow iron K emission lines (K$\alpha$, K$\beta$ and the Compton 
shoulder) are shown as solid black lines. The dual reflector model has 
reflection fractions of $R\sim0.9$ and $R\sim0.5$ for the disk and distant 
components respectively. The data are consistent with equal contributions from 
the disk and distant reflectors, within the errors.} 
\end{figure}

Both reflection components are required 
to model the Suzaku data. If the distant reflector and narrow iron line 
are removed and the spectrum refitted with just the ionized disk emission, 
then the fit statistic is much worse ($\chi^{2}/{\rm dof}=2261/1882$, model 4, Table 3), 
while the disk reflection fraction becomes rather large ($R=1.9$). Alternatively 
if the disk reflector is removed (which includes the emission from the broad 
iron line) and the spectrum is just fitted with the distant reflector 
including the narrow iron line core, then the fit statistic is also worse
($\chi^{2}/{\rm dof}=2027/1882$, model 2, Table 3), while the strength of the distant reflector 
increases to $R=1.5\pm0.2$ in order to compensate for the lack of the disk reflector 
and broad iron line. 

\begin{table*}
\caption{Comparison of iron line and reflection model fits to the mean spectrum. See text for model details.}
  \begin{center}
   \begin{tabular}{llc}
\hline
     Model & Description & $\chi^{2}/dof$ \\
\hline   
1. & Neutral Reflection (pexrav) + Narrow and Broad Iron lines & 1983/1879 \\
2. & Neutral Reflection + Narrow Iron Line only & 2027/1882 \\
3. & Ionized Disk Reflection + Neutral reflection and Narrow Iron line & 1980/1878 \\
4. & Ionized Disk Reflection only & 2261/1882 \\
\hline
     \end{tabular}
  \end{center}
\end{table*}
 
\section{Discussion and Conclusions}

The long Suzaku observation of the bright, Seyfert galaxy MCG\,-5-23-16 has 
revealed all the emission components expected in the X-ray spectrum of an 
obscured Seyfert galaxy in the context of AGN unified schemes 
\citep{Ant93}.  A canonical high energy power-law ($\Gamma=1.9-2.0$) 
continuum is observed, absorbed at low 
energies by Compton-thin matter and rolling over at high energies with a cut-off energy 
of $>170$\,keV. 
At soft X-ray energies a steep unabsorbed soft 
excess is observed, probably a combination of scattered power-law emission 
and photoionized line emission from distant gas. Such a component is 
common in many Seyfert 2 galaxies \citep{Turner97, Sambruna01, Kink02, Pounds05, Bianchi05}. 
Importantly, the excellent coverage of Suzaku at high energies has allowed us to obtain 
extremely accurate constraints on the iron K line profile and the high energy reflection 
component. Without a direct measurement of the reflection hump above 10 keV, 
the amount of reflection present in the spectrum would be uncertain, 
making the determination of both the underlying continuum and the iron K line 
parameters somewhat degenerate.
Due to the broad bandpass and high signal to noise 
of the Suzaku data, both the broad and narrow components of the iron K$\alpha$ line as 
well as the reflection hump are unambiguously detected. The broad line is clearly 
resolved, with a velocity width of $\sigma_{\rm v}=20000$\,km\,s$^{-1}$ 
(or 46000\,km\,s$^{-1}$ FWHM), which requires emission within $\sim100R_{\rm g}$ 
of the putative massive black hole in MCG\,-5-23-16. 

\subsection{The Geometry and Location of the Reprocessing Matter in MCG\,-5-23-16}

The high signal to noise of the Suzaku XIS spectrum, together 
with the simultaneous Chandra/HETG observation, limits
the width of the narrow iron K$\alpha$ line core. The upper limit obtained with Suzaku is
$\sigma_{\rm v}<2200$\,km\,s$^{-1}$. 
For a black hole mass of $5\times10^{7}{\rm M}_{\odot}$ 
\citep{WM86}, this 
corresponds to a distance of $>0.05$\,pc or $>2\times10^{4}R_{\rm g}$. 
At this distance and given the detection of the distant 
reflection component observed in the HXD, 
a likely origin of the line is from fluorescence and scattering off material 
in a Compton-thick molecular torus \citep{Ghis94, Krolik94}. 
Alternatives to the torus could be a bi-conical 
outflow \citep{Elvis2000}, or reflection off the outer radii of a warped disk 
\citep{Nayakshin05}. 

Two zones of Compton thick matter may exist in MCG\,-5-23-16, the 
accretion disk being responsible for the broad iron K$\alpha$ line, while the torus is 
the likely candidate for the distant reprocessor. This was first suggested for 
this source by \citet{Weaver97} on the basis of a two component model fit to the 
ASCA iron line profile. The dual reflection fits presented 
above suggest that the torus is likely Compton-thick (i.e. $N_{\rm H}>10^{24}$\,cm$^{-2}$),  
while the iron abundance in the torus is about half the solar value. 
The likely geometry for MCG\,-5-23-16 is that we are viewing through Compton-thin 
material at the edge of the torus which is Compton-thick at angles along the plane of 
the accretion disk. This picture seems consistent with the inclination 
angle derived from the diskline and disk reflection fits to the broad iron line, 
of about 45\degg. Similarly 
\citet{WR98} also found inclination angles in the range 40--50\degg\ 
for the iron line fits to a sample of 4 Compton-thin Seyfert 2s (including MCG\,-5-23-16). 
Indeed the fact that MCG\,-5-23-16 displays broad 
Paschen $\beta$ in the infra-red \citep{Goodrich94} 
is consistent with viewing the inner BLR through a moderate line of sight column. 

An interesting question is how many of the Compton-thin 
Seyfert 2 galaxies show evidence for a Compton-thick reprocessor? 
The most detailed study to date of the high energy emission 
from Compton-thin Seyfert 2s was in an analysis of Beppo-SAX observations by 
\citet{Risaliti02}. In the large majority of cases in this sample (17 out of 21 objects) 
a significant detection of Compton reflection has been found, while the authors favor a 
distant origin for the reflector from its apparent lack of variability. 
On the other hand the line of sight 
column densities of the Compton-thin Seyferts is too low 
(e.g. $N_{\rm H}<10^{23}$\,cm$^{-2}$) to have sufficient efficiency for Compton 
down--scattering 
high energy photons. Thus the scenario whereby at least two zones of 
absorbing matter exists, 
with the Compton-thick zone out of the direct line of sight, would appear to be 
fairly commonplace in Compton-thin AGN. 

In the case of MCG\,-5-23-16, the column density 
of the line of sight absorber does not appear to significantly change with time; the 
column density measured by Suzaku ($1.65\pm0.03\times10^{22}$\,cm$^{-2}$) lies 
close to the 
mean value of $\sim1.7\times10^{22}$\,cm$^{-2}$ observed by ASCA, Beppo-SAX, 
XMM-Newton and 
Chandra from 1994--2001, with little scatter \citep{Balestra04} 
and it also agrees well with earlier 
measurements \citep{REN02}. 
Thus this absorber is likely to reside at large distances from the black hole (e.g. the torus  
or even the host galaxy) and is likely not to be clumpy. Finally we note that 
observations of Seyfert 2s, where the AGN appears to change from Compton-thick (i.e. 
reflection dominated) to Compton-thin or vice versa, 
also argues for the existence of at least two zones 
of distant absorbing matter \citep{Matt03}. 
While the Compton-thick reprocessor could originate from parsec scale material 
such as the torus, it is possible that 
the Compton-thin X-ray absorber originates from circumnuclear gas or dust on a larger 
(e.g. $\sim100$\,pc) scale within the AGN host galaxy 
\citep{Maiolino95, Malkan98, Matt2000, Guainazzi05}. 

\subsection{The Nature of the Iron K Line Emission}

The lack of variability of the reflection spectrum in MCG\,-5-23-16 on short timescales 
is also consistent with part of the reflection originating in distant matter. As the 
high -- low 
difference spectrum illustrates, the only variable part of the spectrum appears 
to be the intrinsic power-law from the disk/corona. While the distant 
reprocessor is not expected to vary rapidly, the lack of short-term variability 
of the broad line in MCG\,-5-23-16 is curious. The best studied example to date is 
MCG\,-6-30-15, where the relativistic iron line also appears not to vary 
despite strong continuum fluctuations
\citep{VF04}, with gravitational light bending of the continuum photons near 
the black hole event horizon being one possible explanation \citep{MF04}. 
In MCG\,-5-23-16 the iron line emission is much less centrally 
concentrated, with the line profile characterized by an inner radius of about 
$\sim20-40R_{\rm g}$, unlike for MCG\,-6-30-15 where the disk emission 
is likely to extend into $6R_{\rm g}$ or even $1.2R_{\rm g}$ 
\citep{Wilms01, Fabian02}. 
Taking a black hole mass estimate of $5\times10^{7}$M$_{\odot}$  
and if the line emission originates from radii $>20R_{\rm g}$, 
then the reverberation timescale for the broad iron line will be $>10^{4}$\,s, which 
should be observable over the Suzaku observation (of 220\,ks duration). One possibility 
is that the continuum flux variations are not strong enough (about 40\%) in this source 
to produce a detectable difference in the broad iron K line flux. 
Suzaku observations of other bright, variable Seyfert galaxies 
with confirmed broad iron lines will be able to 
address this issue. An early indication is given by the deep (350\,ks) 
Suzaku observation of MCG\,-6-30-15, which despite the rapid continuum flux 
variability, shows no strong variations in the iron line or reflection hump 
with continuum flux \citep{Miniutti06}.

Indeed the inner radius implied from the iron line profile could suggest 
that the inner disk may be missing or truncated in MCG\,-5-23-16. If the accretion 
rate is low enough, the innermost disk can transition to an advective dominated 
accretion flow or ADAF \citep{NY95}. In MCG\,-5-23-16 the 2--10\,keV 
X-ray luminosity is $1.5\times10^{43}$\,erg\,s$^{-1}$. Thus if the 2--10\,keV 
represents $\sim5\%$ of the bolometric output in a typical AGN \citep{Elvis94}, 
then the bolometric luminosity of MCG\,-5-23-16 is of the order 
$\sim3\times10^{44}$\,erg\,s$^{-1}$. For a black hole mass of  
$5\times10^{7}$M$_{\odot}$, the accretion rate of MCG\,-5-23-16 is $\sim5$\% 
of the Eddington rate. Interestingly, this may be close to the transition rate between 
high/soft and low/hard states in Galactic black hole sources, thought to be a 
few percent of Eddington \citep{Maccarone03}. So it may be plausible that the 
inner disk could be truncated in MCG\,-5-23-16 at a few tens of gravitational radii. 
Conversely it is perhaps thought that most X-ray bright Seyfert galaxies 
are more analogous to high state black hole sources, at least based on their 
power density spectra \citep{UM05}.

However an optically-thick disk may extend inwards to the last stable orbit in 
MCG\,-5-23-16. If the emissivity of the X-ray source illuminating the disk 
is fairly flat (e.g. varying as $R^{-2}$), then 
the disk may extend all the way to an inner radius of $6R_{\rm g}$ or even 
$1.2R_{\rm g}$. In the spectral fits, this provides an  
equally acceptable fit to the data. 
A flat emissivity profile 
could occur if the illuminating source is located high above the disk in the 
``lamp-post'' geometry, or if the X-ray emission is dissipated from flares further out on 
the disk. Furthermore it is possible that matter in the innermost disk 
radii is close to fully ionized. The reflection spectrum 
of a highly ionized disk or slab can be largely featureless and resemble that of the input 
continuum, especially if iron becomes fully ionized down to several Thomson depths 
\citep{Nayakshin00, Ballantyne01}. In this case the inner disk radius 
could correspond to a characteristic radius at which the 
disk surface becomes fully ionized. 

An alternative possibility is that the broad iron line 
may not even originate from the accretion disk. 
The emission could originate from reprocessing 
in either a spherical distribution of clouds \citep{GR88} or a 
bi-conical outflow \citep{Sulentic98, Elvis2000}. 
Ultra-fast outflows 
have been claimed in a handful of AGN, on the basis of blue-shifted 
high ionization iron K
absorption lines or edges; e.g. PG 1211+143 \citep{Pounds03b}, PDS 456 \citep{R03}, 
IC\,4329a \citep{Markowitz06}. Indeed the long December 2005 XMM-Newton observation of 
MCG\,-5-23-16 
shows evidence for a variable, blue-shifted iron K$\alpha$ absorption line at 7.8\,keV 
(Braito et al. 2006), which could plausibly arise from such an outflow. 

A broader issue is the apparent lack of broad iron lines detected
in recent XMM-Newton observations of bright Seyfert galaxies. Despite the high quality of 
the observations, sometimes only a narrow line is present \citep{Pounds03a, Schurch03, 
Bianchi04}. In some cases broad residuals are present in the 
iron K band, but the continuum curvature due to a high column warm absorber and a 
red-wing of a relativistic iron line are difficult to distinguish 
\citep{R04, Pounds04, Turner05}. In the example of MCG\,-5-23-16, 
with Suzaku it is clear that a broad iron line is present. 
However in a much shorter observation of a weaker source such a line may not 
be detected, or its modeling could be ambiguous where there is no bandpass 
above 10\,keV. As discussed above, the disk lines may be weaker than expected if the 
inner accretion disk is close to fully ionized; this will be dependent both on the 
illuminating continuum and the X-ray emission geometry \citep{Nayakshin00}.
Also if the iron abundance 
is less than Solar (as measured in MCG\,-5-23-16), the broad line will be weaker. 


In AGN X-ray spectra where complex or even partial covering absorption
appears to be present (e.g. NGC\,3516 \cite{Turner05}, NGC\,4051 \cite{Pounds04}), 
one of the major problems in assessing the contribution of 
the broad iron line is in the uncertainty in determining the underlying continuum. 
Suzaku presents the best capabilities of the current X-ray missions for resolving this 
issue, by revealing the true shape and level of the underlying continuum and 
reflection component through 
broadband observations with coverage above 10\,keV, while also achieving high signal 
to noise (at least equal to XMM-Newton) in the iron K band. Encouragingly there appears to 
be a large number of hard X-ray selected AGN emerging from the Swift/BAT All Sky
Survey \citep{Markwardt05}, which could eventually number as a many as $\sim200$ 
Type I and Type 2 AGN at a limiting flux level of $\sim10^{-11}$\,erg\,cm$^{-2}$\,s$^{-1}$. 
Most of these AGN will be bright enough for detailed study 
both in the iron K bandpass and above 10\,keV with Suzaku. 
These sources can form the basis of our 
future understanding of the iron line and Compton reflection associated with 
the innermost regions of AGN and for testing our understanding of Unified models and 
AGN evolution. 









\begin{thebibliography}{}

\bibitem[Antonucci(1993)]{Ant93}
Antonucci, A. 1993, ARA\&A, 31, 473

\bibitem[Ballantyne et al.(2001)]{Ballantyne01}
Ballantyne, D.R., Ross, R.R., \& Fabian, A.C. 2001, MNRAS, 327, 10

\bibitem[Balucinska-Church \& McCammon(1992)]{BM92} 
Balucinska-Church, M. \& McCammon, D. 1992, ApJ, 400, 699

\bibitem[Bassani et al.(2006)]{Bassani06}
Bassani, L. et al. 2006, ApJ, 636, L65

\bibitem[Balestra et al.(2004)]{Balestra04}
Balestra, I., Bianchi, S., \& Matt, G. 2004, A\&A, 415, 437  

\bibitem[Bianchi et al.(2004)]{Bianchi04}
Bianchi, S., Matt, G., Balestra, I., Guainazzi, M., \& Perola, G.C. 2004, 
A\&A, 422, 65

\bibitem[Bianchi et al.(2005)]{Bianchi05} 
Bianchi, S., Miniutti, G., Fabian, A. C., \& Iwasawa, K. 2005, MNRAS, 360, 380

\bibitem[Boldt(1987)]{Boldt87}
Boldt, E. 1987, Phys. Rep., 146, 215

\bibitem[Churazov et al.(2006)]{Chur06}
Churazov, E. et al., 2006, A\&A, in press

\bibitem[Dewangan et al.(2003)]{Dewangan03} 
Dewangan, G.C., Griffiths, R.E., \& Schurch, N.J. 2003, ApJ, 592, 52

\bibitem[Dickey \& Lockman(1990)]{Dickey90} 
Dickey, J.M., \& Lockman, F.J. 1990, ARA\&A, 28, 215

\bibitem[Elvis et al.(1994)]{Elvis94}
Elvis, M., et al., 1994, ApJS, 95, 1

\bibitem[Elvis(2000)]{Elvis2000}
Elvis, M. ApJ, 2000, 545, 63

\bibitem[Fabian et al.(1989)]{Fabian89}
Fabian, A.C., Rees, M.J., Stella, L., \&  White, N.E. 1989, MNRAS, 238, 729

\bibitem[Fabian et al.(2002)]{Fabian02}
Fabian, A.C., et al., 2002, MNRAS, 335, L1


\bibitem[George \& Fabian(1991)]{George91}
George, I.M., \& Fabian, A.C. 1991, MNRAS, 249, 352

\bibitem[Ghisellini et al.(1994)]{Ghis94}
Ghisellini, G., Haardt, F., \& Matt, G. 1994, MNRAS, 267, 743

\bibitem[Goodrich(1994)]{Goodrich94}
Goodrich, R.W., Veilleux, S., \& Hill, G.J. 1994, 422, 521

\bibitem[Gruber et al.(1999)]{Gruber99}
Gruber, D.E., Matteson, J.L., Peterson, L.E., \& Jung, G.V. 1999,
ApJ, 520, 124 

\bibitem[Guainazzi et al.(2005)]{Guainazzi05} 
Guainazzi, M., Matt, G., Perola, G.C. 2005, A\&A, 444, 119

\bibitem[Guilbert \& Rees(1988)]{GR88}
Guilbert, P.W. \& Rees, M.J. 1988, MNRAS, 233, 475


\bibitem[Kinkhabwala  et al.(2002)]{Kink02} 
Kinkhabwala, A. et al., 2002, ApJ, 575, 732

\bibitem[Kokubun et al.(2006)]{Kokubun06}
Kokubun, M. et al., 2006, PASJ, this issue

\bibitem[Koyama et al.(2006)]{Koyama06}
Koyama, K. et al., 2006, PASJ, this issue

\bibitem[Krolik et al.(1994)]{Krolik94}
Krolik, J.H., Madau, P., \& Zycki, P.T. 1994, ApJ, 266, 653

\bibitem[Kushino et al.(2002)]{Kushino02}
Kushino, A., Ishisaki, Y., Morita, U., Yamasaki, N.Y., Ishida, M., Ohashi, T., 
\& Ueda, Y. 2002, PASJ, 54, 327

\bibitem[Laor(1991)]{Laor91}
Laor, A. 1991, ApJ, 376, L90

\bibitem[Lightman \& White(1988)]{LW88}
Lightman, A.P., \& White, T.R. 1988, ApJ, 335, L57

\bibitem[Maccarone(2003)]{Maccarone03}
Maccarone, T.J. 2003, A\&A, 409, 697

\bibitem[Magdziarz \& Zdziarski(1995)]{MZ95}
Magdziarz, P., \& Zdziarski, A. 1995, MNRAS, 273, 837

\bibitem[Maiolino et al.(1995)]{Maiolino95}
Maiolino, R., Ruiz, M., Rieke, G.H., Keller, L.D. 1995, ApJ, 446, 561

\bibitem[Malkan et al.(1998)]{Malkan98} 
Malkan, M.A., Gorjian, V., Tam, R. 1998, ApJS, 117,25

\bibitem[Markowitz \& Edelson(2004)]{ME04}
Markowitz, A.. \& Edelson, R. 2004, ApJ, 617, 939

\bibitem[Markowitz et al.(2006)]{Markowitz06}
Markowitz, A., Reeves, J.N., \& Braito, V. 2006, ApJ, 646, 783

\bibitem[Markwardt et al.(2005)]{Markwardt05}
Markwardt, C.B., Tueller, J., Skinner, G.K., Gehrels, N., Barthelmy, S.D., 
\& Mushotzky, R.F. 2005, ApJ, 633, L77

\bibitem[Matt(2000)]{Matt2000} 
Matt, G. 2000, A\&A, 355, 31L
 
\bibitem[Matt(2002)]{Matt02}
Matt, G. 2002, MNRAS, 337, 147

\bibitem[Matt et al.(2003)]{Matt03} 
Matt, G., Guainazzi, M., Maiolino, R., 2003 MNRAS, 342, 422

\bibitem[Mattson \& Weaver(2004)]{MW04}
Mattson, B.J., \& Weaver, K.A. 2004, ApJ, 601, 771

\bibitem[Miniutti \& Fabian(2004)]{MF04}
Miniutti, G. \& Fabian, A.C. 2004, MNRAS, 349, 1435

\bibitem[Miniutti et al.(2006)]{Miniutti06}
Miniutti, G. et al., 2006, PASJ, this issue

\bibitem[Mushotzky(1982)]{Mushy82}
Mushotzky, R. 1982, ApJ, 256, 92

\bibitem[Mitsuda et al.(2006)]{Mitsuda06}
Mitsuda, K. et al., 2006, PASJ, this issue

\bibitem[Nandra \& Pounds(1994)]{Nandra94}
Nandra, K., \& Pounds, K.A. 1994, MNRAS, 268, 405

\bibitem[Nandra et al.(1997)]{Nandra97}
Nandra, K., George, I.M., Mushotzky, R.F., Turner, T.J., \& Yaqoob, T. 
1997, ApJ, 477, 602

\bibitem[Narayan \& Yi(1995)]{NY95}
Narayan, R. \& Yi, I. 1995, ApJ, 452, 710

\bibitem[Nayakshin et al.(2000)]{Nayakshin00}
Nayakshin, S., Kazanas, D., \& Kallman, T.R. 2000, ApJ, 537, 833

\bibitem[Nayakshin(2005)]{Nayakshin05}
Nayakshin, S. 2005, MNRAS, 359, 545

\bibitem[Page et al.(2004)]{Page04}
Page, K.L., O'Brien, P.T., Reeves, J.N., \& Turner, M.J.L. 2004, 
MNRAS, 347, 316

\bibitem[Perola et al.(2002)]{Perola02}
Perola, G.C., Matt, G., Cappi, M., Fiore, F., Guainazzi, M., Maraschi, L., 
Petrucci, P.O., \& Piro, L. 2002, A\&A, 389, 802 

\bibitem[Pounds et al.(1990)]{Pounds1990}
Pounds, K.A., Nandra, K., Stewart, G.C., George, I.M.,
\& Fabian, A.C. 1990, Nature, 344, 132

\bibitem[Pounds et al.(2003a)]{Pounds03a}
Pounds, K.A., Reeves, J.N., Page, K.L., Edelson, R., Matt, G., 
\& Perola, G.C. 2003a, MNRAS, 341, 953

\bibitem[Pounds et al.(2003b)]{Pounds03b}
Pounds, K.A., Reeves, J.N., King, A.R., Page, K.L., O'Brien, P.T., 
\& Turner, M.J.L. 2003b, MNRAS, 345, 705

\bibitem[Pounds et al.(2004)]{Pounds04}
Pounds, K.A., Reeves, J.N., King, A.R., \& Page, K.L. 2004, MNRAS, 350, 10

\bibitem[Pounds \& Page(2005)]{Pounds05}  
Pounds, K.A., \& Page, K.L. 2005, MNRAS, 360, 1123


\bibitem[Reeves et al.(2003)]{R03} 
Reeves, J.N., O'Brien, P.T., \& Ward, M. 2003, ApJ, 593, L65

\bibitem[Reeves et al.(2004)]{R04} 
Reeves, J.N., Nandra, K., George, I.M., Pounds, K.A., Turner, T.J., 
\& Yaqoob, T. 2004, 602, 648


\bibitem[Reynolds(1997)]{Reynolds97}
Reynolds, C.S. 1997, MNRAS, 286, 513

\bibitem[Risaliti(2002)]{Risaliti02}
Risaliti, G. 2002, A\&A, 386, 379

\bibitem[Risaliti et al.(2002))]{REN02}
Risaliti, G., Elvis, M., \& Nicastro, F. 2002, ApJ, 571, 234

\bibitem[Ross \& Fabian(2005))]{RF05}
Ross, R.R. \& Fabian, A.C. 2005, MNRAS, 358, 211

\bibitem[Sambruna et al.(2001)]{Sambruna01}
Sambruna, R.M., et al., 2001, ApJ, 546, L13

\bibitem[Schurch et al.(2003)]{Schurch03}
Schurch, N., Warwick, R.S., Griffiths, R.S., \& Sembay, S. 2003, 
MNRAS, 345, 423

\bibitem[Sulentic et al.(1998)]{Sulentic98}
Sulentic, J.W., Marziani, P., Zwitter, T., Calvani, M., \& Dultzin-Hacyan, D. 
1998, ApJ, 501, 54

\bibitem[Takahashi et al.(2006)]{Takahashi06}
Takahashi, T. et al., 2006, PASJ, this issue

\bibitem[Tanaka et al.(1995)]{Tanaka95}
Tanaka, Y. et al. 1995, Nature, 375, 659


\bibitem[Turner et al.(1997)]{Turner97}
Turner, T.J., George, I.M., Nandra, K., \& Mushotzky, R.F. 1997, ApJ, 488, 164

\bibitem[Turner et al.(2005)]{Turner05}
Turner, T.J., Kraemer, S.B., George, I.M., Reeves, J.N., \& Bottorff, M C. 
2005, 618, 155

\bibitem[Uttley \& McHardy(2005)]{UM05}
Uttley, P. \& McHardy, I.M. 2005, MNRAS, 363, 586

\bibitem[Vaughan \& Fabian(2004)]{VF04}
Vaughan, S. \& Fabian, A.C. 2004, MNRAS, 348, 1415

\bibitem[Veron et al.(1980)]{Veron80}
Veron, P., Lindblad, P.O., Zuiderwijk, E.J., Veron, M.P., \& Adam, G. 
1980, A\&A, 87, 245


\bibitem[Wandel \& Mushotzky(1986)]{WM86}
Wandel, A., \&  Mushotzky, R.F. 1986, ApJ, 306, L61

\bibitem[Weaver \& Reynolds(1998)]{WR98}
Weaver, K.A., \& Reynolds, C.S. 1998, ApJ, 503, L39

\bibitem[Weaver et al.(1997)]{Weaver97}
Weaver, K.A., Yaqoob, T., Mushotzky, R.F., Nousek, J., Hayashi, I., 
\& Koyama, K. 1997, ApJ, 474, 675

\bibitem[Wegner et al.(2003)]{Wegner03}
Wegner, G., et al., 2003, AJ, 126, 2268

\bibitem[Wilms et al.(2000)]{Wilms00}
Wilms, J., Allen, A., McCray, R. 2000, ApJ, 542, 914

\bibitem[Wilms et al.(2001)]{Wilms01}
Wilms, J., Reynolds, C.S., Begelman, M.C., Reeves, J., Molendi, S.,
Staubert, R., \& Kendziorra, E. 2001, MNRAS, 328, L27

\bibitem[Yaqoob \& Padmanabhan(2004)]{YP04}
Yaqoob, T., \& Padmanabhan, U. 2004, ApJ, 604, 63

\end{thebibliography}
\end{document}